\documentclass[preprintnumbers,prb]{revtex4}

\usepackage[]{graphicx}

\begin{document} 
\title{Fluctuations in models of biological macroevolution} 
\author{Per Arne Rikvold}\email{rikvold@csit.fsu.edu}
\affiliation{
School of Computational Science, Center for Materials Research and
Technology,\\ 
National High Magnetic Field Laboratory, and Department of Physics\\
Florida State University, Tallahassee, FL 32306, USA}

\begin{abstract}
Fluctuations in diversity and extinction sizes are discussed and
compared for two different, individual-based models of biological
coevolution. Both models display power-law distributions for
various quantities of evolutionary interest, such as
the lifetimes of individual species, the quiet periods between
evolutionary upheavals larger than a given cutoff, and the sizes
of extinction events. Time series of the diversity and 
measures of the size of extinctions give rise to flicker
noise. Surprisingly, the power-law behaviors of the probability
densities of quiet periods in the two models differ, while the
distributions of the lifetimes of individual species are the same. 
\end{abstract}



\maketitle 

\section{INTRODUCTION}
\label{sec:int}  

Biological evolution presents many problems concerning
highly nonlinear, nonequilibrium systems of interacting entities
that are well suited for study by methods from statistical
mechanics. An excellent review of this rapidly growing field is
found in Ref.~\onlinecite{DROS01}. Among these problems are those
concerned with {\it coevolution\/}: the simultaneous evolution of
many species, whose mutual interactions 
produce a constantly changing {\it fitness
landscape\/}. Throughout this process, new species are created and
old species go extinct. 
A question that is still debated in evolutionary biology is whether
evolution on the macroscale proceeds gradually or in `fits and starts,'
as suggested by Eldredge and Gould.\cite{ELDR72,GOUL77,GOUL93}
In the latter mode, known as {\it punctuated equilibria\/}, species and
communities appear to remain largely unchanged for long periods of time,
interrupted by brief (on a geological timescale) periods of mass
extinctions and rapid change. 

A coevolution process involves a
large range of timescales, from the ecologically relevant scales of
a few generations, to geological scales of millions or billions of
generations. Traditionally, models of macroevolution have been
constructed on a highly coarse-grained timescale. (The best-known
such model to physicists is probably the Bak-Sneppen
model.\cite{BAK93}) However, the long-time dynamics of the
evolution is clearly driven by ecological processes, mutations, and
selection at comparatively short timescales. As a result, in recent
years several new models have been proposed that are designed to span
the disparate ecological and evolutionary timescales. These models
include the Webworld model,\cite{CALD98,DROS01B,DROS04} 
the Tangled-nature model,\cite{HALL02,CHRI02} and simplified versions of
the latter.\cite{RIKV03,ZIA04,SEVI05} 
In this paper I discuss and compare some of the properties of 
fluctuations in two simplified coevolution models: 
the model introduced in Ref.~\onlinecite{RIKV03} 
and a different model that I am currently developing.\cite{RIKV05}

The rest of this paper is organized as follows. In Sec.~\ref{sec:mod} 
I introduce the two models, in Sec.~\ref{sec:res} I compare and discuss
numerical results from large-scale 
Monte Carlo simulations for the different 
models, and in Sec.~\ref{sec:conc} I present a summary and conclusions.

\section{MODELS}
\label{sec:mod}  

Both of the models studied here consider haploid species whose genome is 
represented by a bit
string of length $L$, so that there are a total of $2^L$ potential
species, labeled by an index $I \in [0,2^L-1]$. 
At the end of each generation, an individual of species $I$ can 
with probability $P_I$ give rise to $F$ offspring and then die, 
or with probability $(1-P_I)$ it dies without offspring. 
The fecundity $F$ is assumed fixed and will be chosen appropriately for
each model as discussed below. 
During its birth, an offspring individual may undergo mutation with a
small probability $\mu/L$ per bit in its genome. During a mutation
event, a bit in the genome is flipped, and the mutated offspring 
individual is considered as belonging to a different species than 
its parent. 

The reproduction probability $P_I(t)$ of species $I$ in generation $t$ 
depends on the interactions
of individuals of that species with other species, $J$, that are present in
the community with nonzero populations $n_J$, as well as possibly on its
ability to utilize an external resource, $R$. It is given by the
convenient nonlinear form,
\begin{equation}
P_I(t) = \frac{1}{1 + \exp[\Phi_I(R,\{n_J\})]} \;,
\label{eq-PI}
\end{equation}
which resembles the acceptance probability in
a Monte Carlo simulation using Glauber transition rates.\cite{LAND00}
The function $\Phi_I$ depends linearly on the set of all nonzero
populations in generation $t$, $\{n_J(t)\}$, and possibly also on the
external resource $R$. When $\Phi_I$ is large and negative, the reproduction
probability is near unity, while large positive values lead to a low
reproduction probability. The coupling to the $n_J$ is effected via the
interaction matrix $\bf M$, whose elements $M_{IJ}$ are 
chosen randomly (with
restrictions discussed below for the individual models) from a uniform
distribution on $[-1,+1]$. 
If $M_{IJ} > 0$ and $M_{JI} < 0$, species $I$ is a predator and $J$ its
prey, or {\it vice versa\/}, while $M_{IJ}$ and $M_{JI}$ both positive
denotes mutualism or symbiosis, and both negative denote competition. 
Once $\bf M$ is chosen at the beginning of
a simulation, it remains fixed (``quenched randomness"). These interactions 
(together with other, model-specific parameters) determine whether or
not a species will have a sufficient reproduction 
probability to be successful
{\it in a particular community\/} of other species. Typically, only a
small subset of species have nonzero populations at
any one time, forming a community. 

\noindent
{\bf Model A:}\\
In Model A, which was introduced and studied in
Refs.~\onlinecite{RIKV03,ZIA04,SEVI05}, $\Phi_I$ takes the form
\begin{equation}
\Phi_I^{\rm A}(\{n_J(t)\}) 
= - \sum_J M_{IJ} n_J(t)/N_{\rm tot}(t) + N_{\rm tot}(t)/N_0
\;,
\label{eq-PhiA}
\end{equation}
where $N_{\rm tot}(t) = \sum_J n_J(t)$ is the total population at $t$. 
In this model, the Verhulst factor $N_0$ prevents the population from
indefinite growth and can be seen as representing an environmental
``carrying capacity."\cite{VERH1838} The interaction matrix has elements
that are randomly distributed over $[-1,+1]$, except that $M_{II}=0$ to
focus on the effects of interspecies interactions. 
Here, Model A will be simulated with the following parameter values:
$F=4$ (see below), $N_0 = 2000$, and $\mu = 10^{-3}$.

\noindent
{\bf Model B:}\\
The new Model B is a predator/prey model. It is somewhat more realistic 
than Model A in that the population is
maintained by a subset of the  
species that can utilize the external resource $R$ 
(primary producers, or autotrophs). Other species 
can maintain themselves only by predation on one or more
of the producer species (consumers, or heterotrophs). 
These modifications lead to the restrictions that
the off-diagonal part of $\bf M$ must be antisymmetric, and further that
if $I$ is a producer and $J$ a consumer, then $I$ must be the prey
of $J$, so that $M_{IJ} < 0$. 
To avoid runaway population growth,
the diagonal elements are chosen randomly on $[-1,0)$. Such negative self
interactions are commonly used in population dynamics
models.\cite{MURR89} 
The specific form of $\Phi_I$ in this model is 
\begin{equation}
\Phi_I^{\rm B}(R,\{n_J(t)\}) 
= b_I - R \eta_I /N_{\rm tot}(t) - \sum_J M_{IJ} n_J(t)/N_{\rm tot}(t)
\;,
\label{eq-PhiB}
\end{equation}
where $b_I > 0$ can be seen as a ``cost of reproduction," and $\eta_I >
0$ is the coupling to the external resource. The latter is positive for
producers and zero for consumers. In this implementation of the model,
both $b_I$ and the nonzero $\eta_I$ are chosen uniformly from $(0,1]$
at the beginning of the simulation and kept fixed thereafter. 
Here, Model B will be simulated with the following parameter values:
$F=2$ (see below), $R = 2000$, and $\mu = 10^{-3}$. 
Only 5\% of the potential species are producers (i.e., have $\eta_I>0$), 
and in order to obtain food webs with more realistic
connectivity, 90\% of the $M_{IJ}/M_{JI}$ pairs are 
randomly chosen to be zero, giving a 
{\it connectance\/}\cite{GARD70,MAY72,YODZ80,DUNN02,GARL03,GARL04} 
of 10\%, while the elements of the nonzero pairs are chosen on
$[-1,+1]$ as described above. 

Both models are sufficiently simple that the populations and stabilities
of fixed-point communities can be obtained analytically for 
zero mutation rate. These analytical results are extensively
compared with simulations in Refs.~\onlinecite{RIKV03,ZIA04},
and~\onlinecite{RIKV05}. For
the purposes of the present study I just mention that the analytical
stability results yield intervals for the fecundity $F$, 
inside which it is ensured that a fixed-point community in the
absence of mutations is stable toward small deviations
in all directions. The values of $F$ used here are chosen to yield such
stable fixed-point communities for both models. 
\begin{figure}[t]
\begin{center}
\begin{tabular}{c}
\includegraphics[angle=0,width=.45\textwidth]{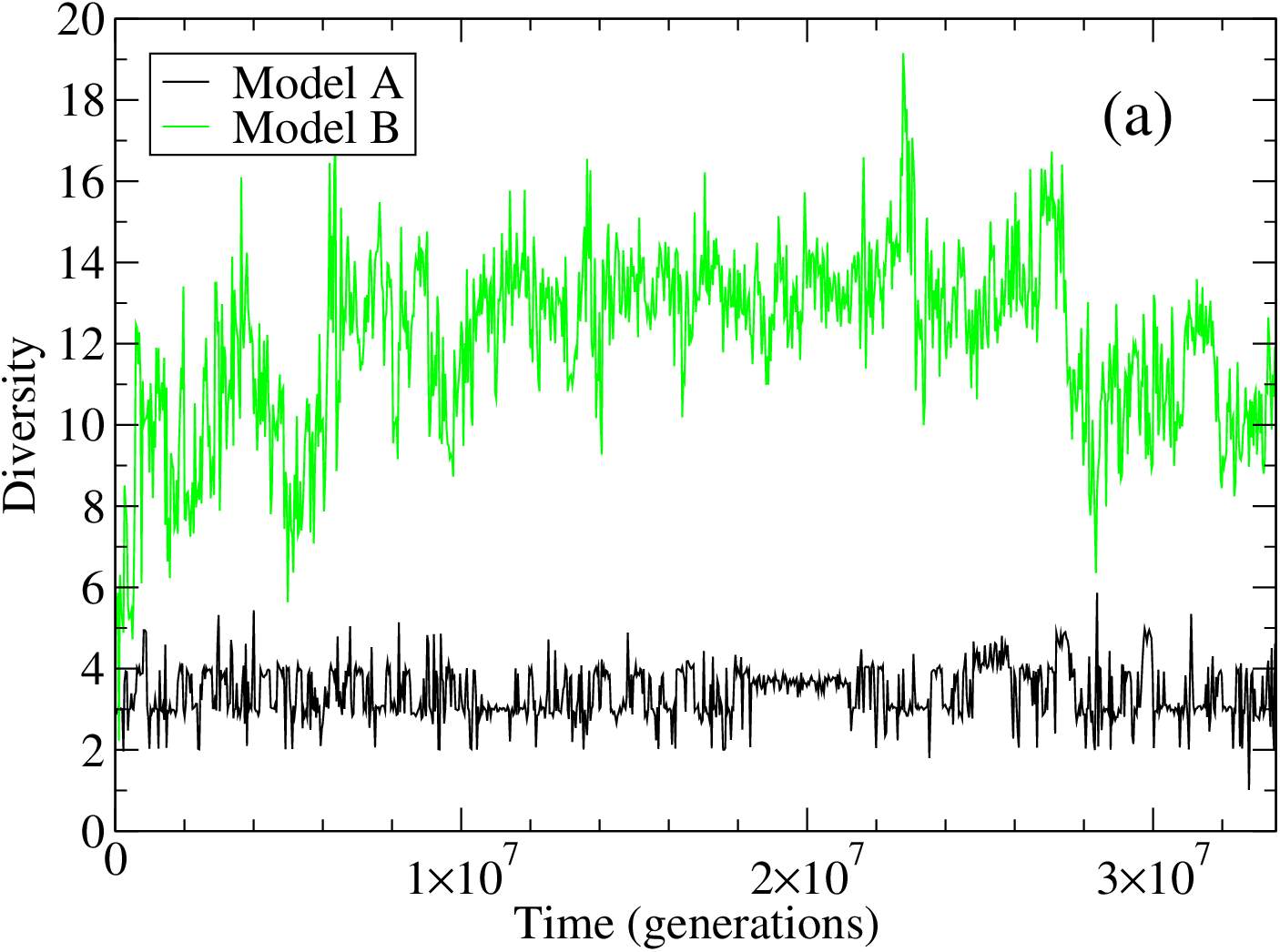}
\hspace{0.5truecm}
\includegraphics[angle=0,width=.45\textwidth]{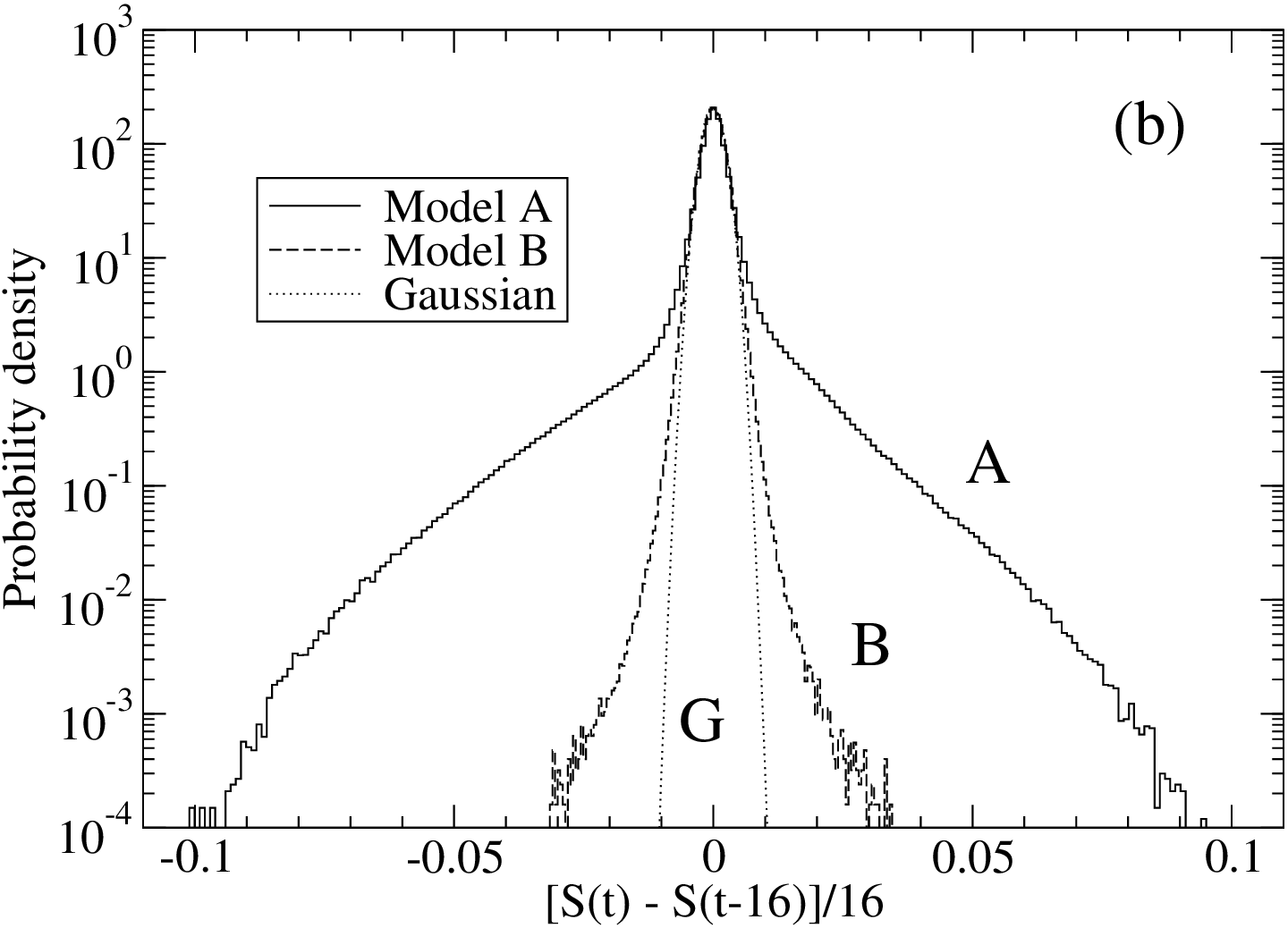}
\end{tabular}
\end{center}
\caption[]
{\label{fig:D}
{\bf (a):}
Typical diversity time series of $T = 2^{25} = 33\,554\,432$ generations 
from Model A (black) and Model B (gray, green online). 
To facilitate printing,
the data are sampled only every $2^{15} = 32\,768$ generations, 
corresponding roughly to the graphical resolution of the
figure. 
While this somewhat reduces the apparent range of the fluctuations,
the general shape of the time series is preserved. 
{\bf (b):}
Histograms representing the probability density 
of the logarithmic derivative of the diversity, $dS(t)/dt$. The data were 
averaged over 16 generations in each run, and then averaged over 
16 independent runs for Model A (solid, marked A) and 12 runs for Model B
(dashed, marked B).  
The central parts of both histograms are well fitted by {\it the same\/}
Gaussian distribution (dotted, marked G). 
\vspace{0.7truecm}
}
\end{figure}

My focus in this paper is to compare the fluctuations and long-time
correlations in the two models 
for quantities of significance in ecology and evolutionary
biology. One such quantity is the {\it diversity\/} 
of a community. This quantity is defined in many ways in the
literature, most simply as the total number of species that exist with
nonzero populations at a given time (also known as the
{\it species richness\/}).\cite{KREB89} Here I define it as 
$D(t) = \exp\left[S\left(\{n_I(t)\}\right)\right]$, where $S$ is the
information-theoretical entropy \cite{SHAN48,SHAN49},
\begin{equation}
S\left( \{ n_I(t) \} \right)
=
- \sum_{\{I | \rho_I(t) > 0 \}} \rho_I(t) \ln \rho_I(t)
\label{eq:S}
\end{equation}
with $\rho_I(t) = n_I(t) / N_{\rm tot}(t)$. This definition of
diversity has the advantage that it is weighted away from species with
very small populations, which in these models are likely to be
short-lived, unsuccessful mutants. In ecology it is known as the
Shannon-Wiener index.\cite{KREB89} It is expected to exhibit a low
noise level during periods with little apparent evolutionary turnover 
(periods of {\it stasis\/}\cite{ELDR72,GOUL77,GOUL93}), while
it should fluctuate much more strongly during highly active periods,
such as mass extinctions or the emergence of new major species. 
Other quantities that can be monitored, and are 
expected to give similar information, are the number of species that go
extinct in a single generation, either by itself, or weighted by the
maximum population attained by the species that go extinct.

\section{NUMERICAL RESULTS AND ANALYSIS}
\label{sec:res}  

\begin{figure}[t]
\begin{center}
\begin{tabular}{c}
\includegraphics[angle=0,width=.45\textwidth]{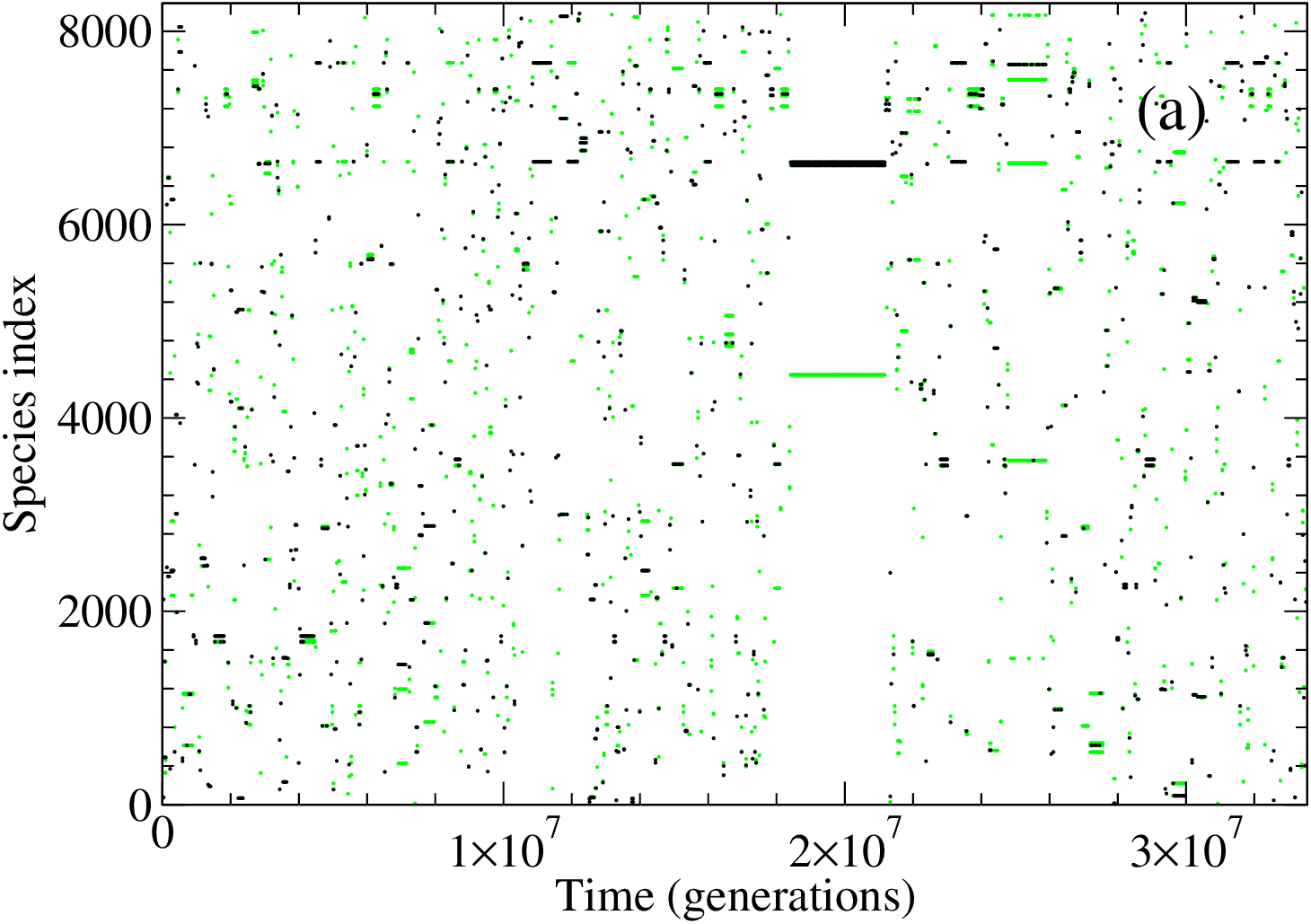}
\hspace{0.5truecm}
\includegraphics[angle=0,width=.45\textwidth]{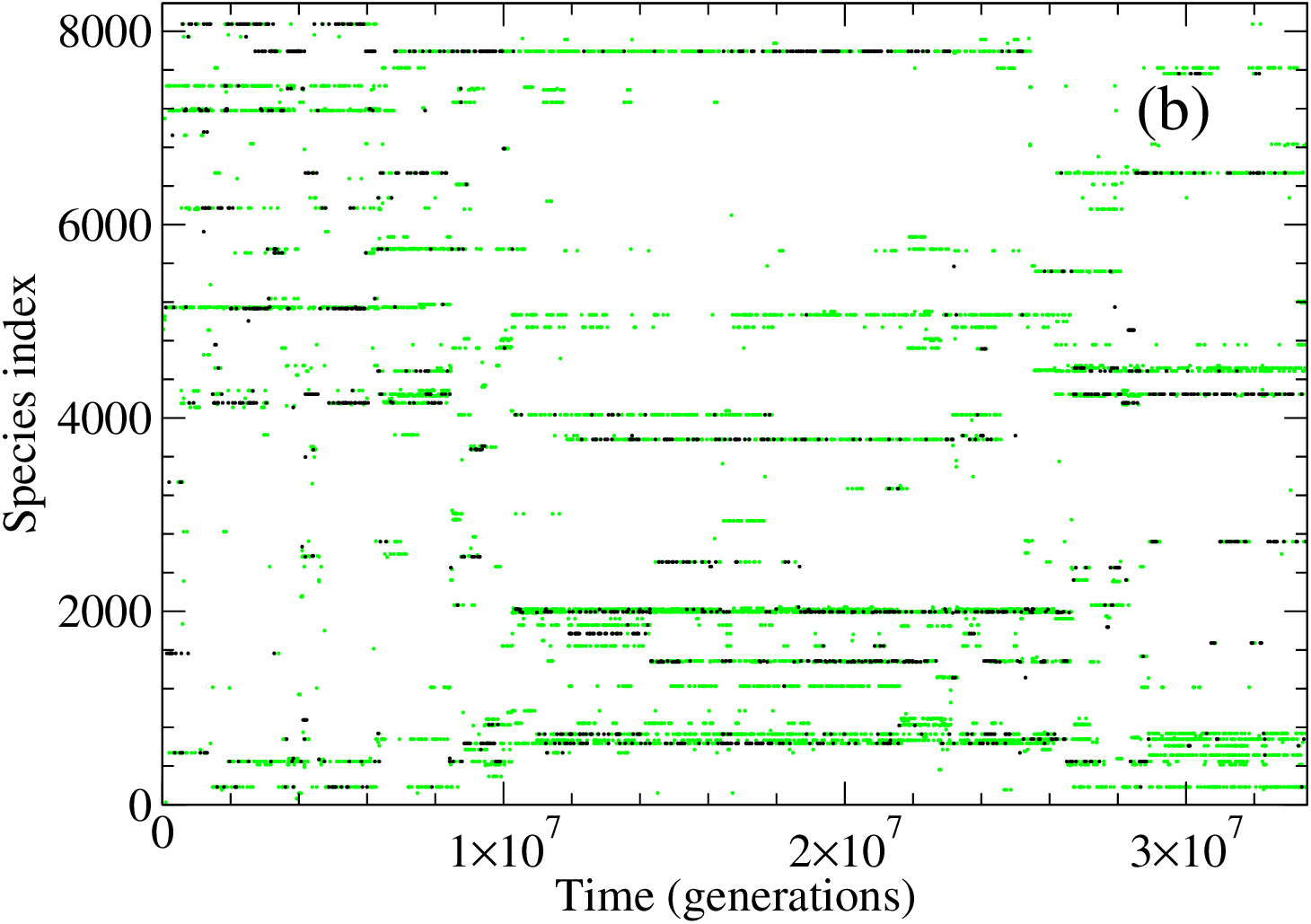}
\end{tabular}
\end{center}
\caption[example]
{\label{fig:spec}
Graphic representation of the community structure vs time,
showing only species with populations between 101 and 1000 (gray,
green online) and above 1000 (black). See discussion in the text. 
{\bf (a):}
Model A. 
Data for the same simulation run illustrated in 
Fig.~\protect\ref{fig:D}(a). 
{\bf (b):}
Model B. 
Data for the same simulation run illustrated in 
Fig.~\protect\ref{fig:D}(a). For ease of plotting, only every
other data point is shown.
}
\end{figure}

\begin{figure}[t]
\begin{center}
\begin{tabular}{c}
\includegraphics[angle=0,width=.45\textwidth]{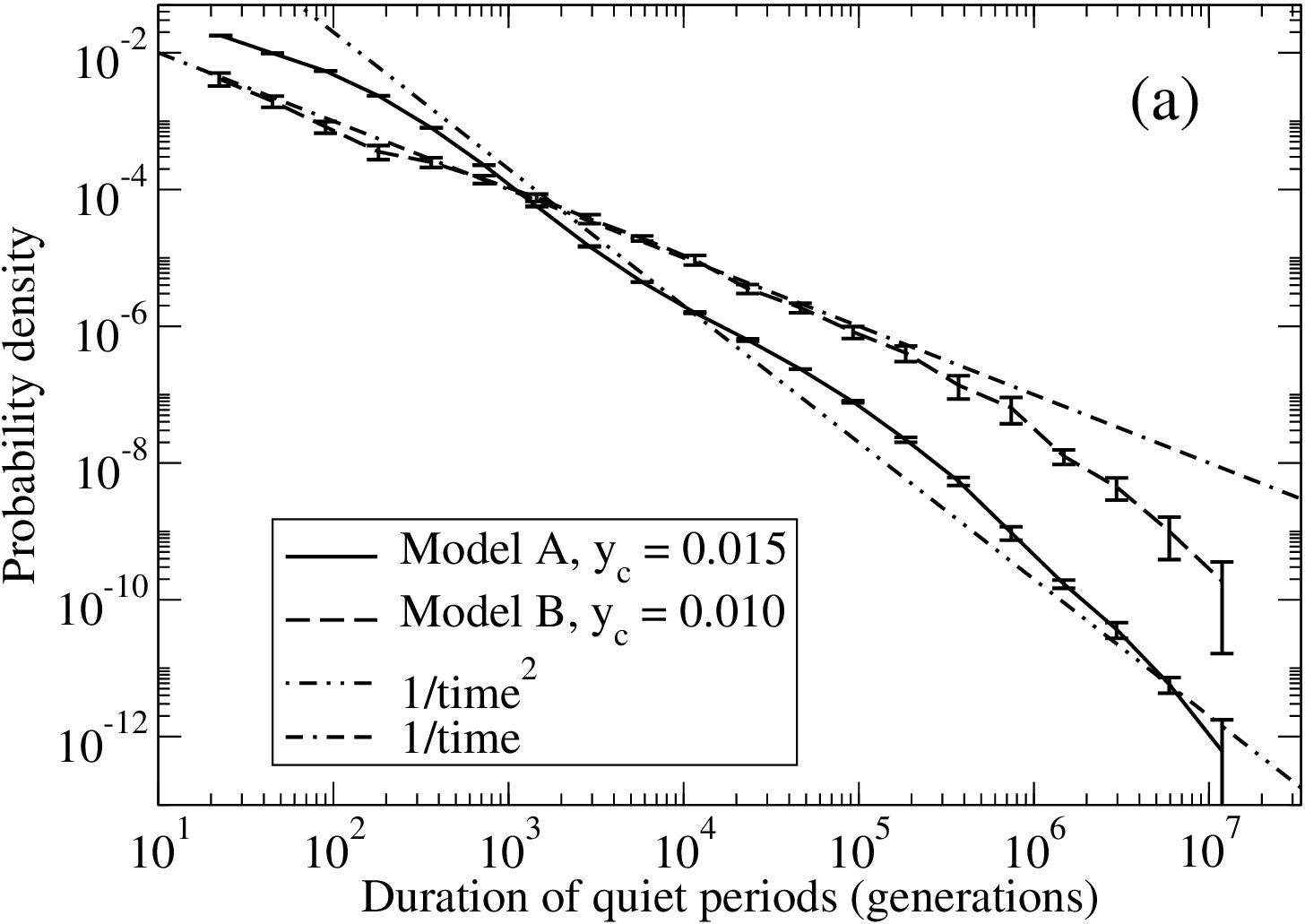}
\hspace{0.5truecm}
\includegraphics[angle=0,width=.45\textwidth]{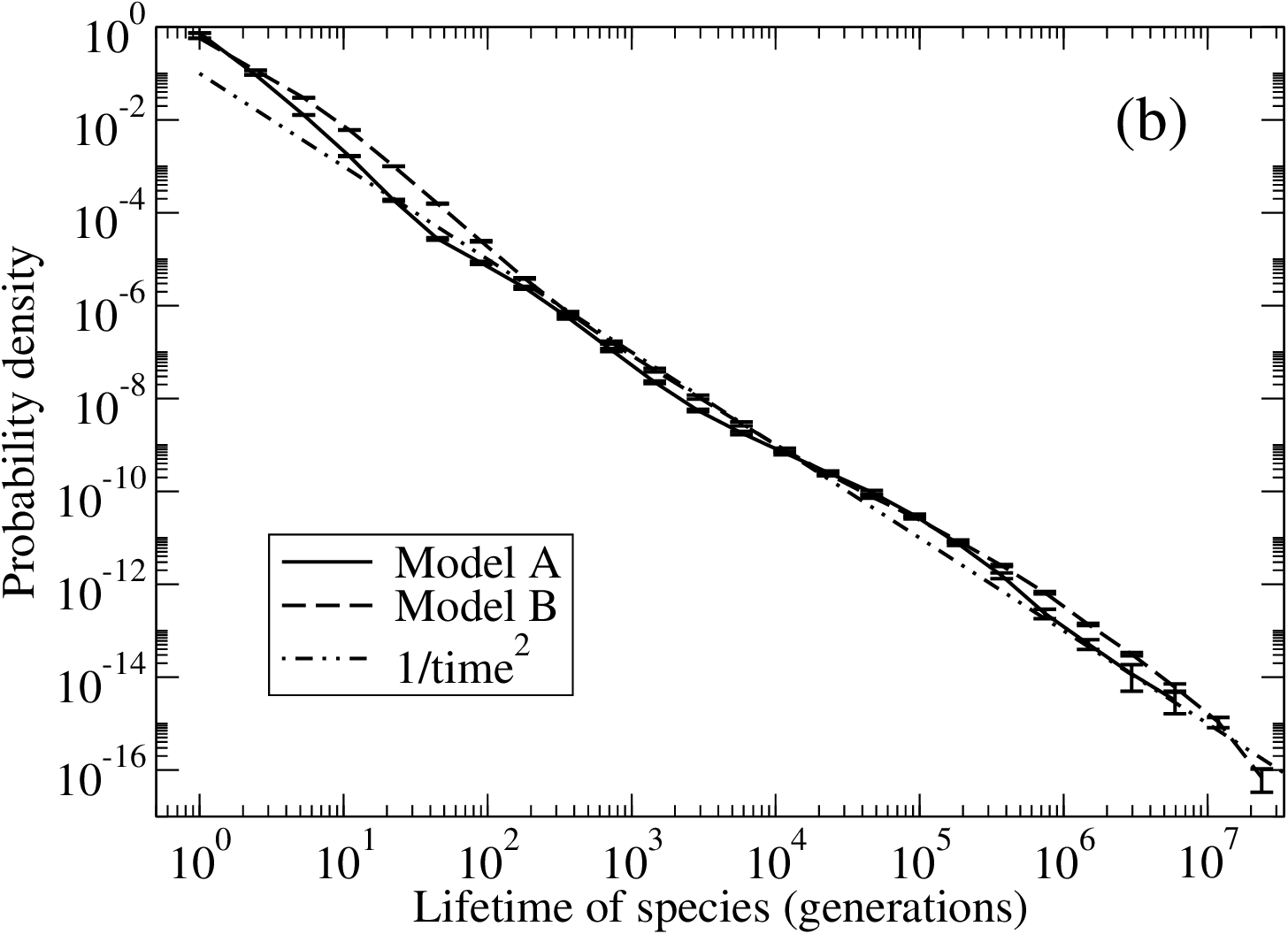}
\end{tabular}
\end{center}
\caption[example]
{\label{fig:dur}
{\bf (a):}
Log-log plot of histograms representing the probability density  
of the durations of quiet periods, estimated as the periods
between times when $y = |dS(t)/dt|$ (averaged over 16 generations)
exceeds a cutoff $y_c$. 
For Model A (solid) and Model B (dashed). 
Results for Model A were averaged over 16 runs, and for Model B
over 12 runs. The error bars are standard errors, based on the spread
between runs. The two dot-dashed straight lines represent 
$1/t^2$ and $1/t$ power laws, respectively. 
{\bf (b):}
Log-log plot of histograms representing the probability density  
of the lifetimes of individual species for Model A (solid) 
and Model B (dashed). 
Results for Model A were averaged over 8 runs, and for Model B
over 12 runs. The error bars were obtained as in part (a). 
The dot-dashed straight line represents a $1/t^2$ power law.
}
\end{figure}
Both models were simulated for $T = 2^{25} = 33\,554\,432$
generations, and several of their statistical properties were compared.
Figure~\ref{fig:D}(a) contains time series for representative
simulation runs for both models. The diversity of Model A
is significantly less than for Model B, but both models show periods
during which the relative fluctuations are modest (quiet periods),
separated by shorter periods during which the diversity changes
strongly. Histograms of the logarithmic derivative of the diversity, 
$d S(t) / dt$, are shown in Fig.~\ref{fig:D}(b). 
In both models the probability densities have Gaussian central peaks of
very nearly the same width, with ``heavy wings." While the Gaussian
peak represents fluctuations due to the stochastic population 
dynamics\cite{ZIA04} 
and the creation and extinction of small populations of unsuccessful
mutants, the wings represent large changes in the composition of the
communities. This can be seen clearly by comparing the timing of the
diversity fluctuations in Fig.~\ref{fig:D} with Fig.~\ref{fig:spec}, 
which shows the
populations of individual species versus time for the same simulation
runs shown in Fig.~\ref{fig:D}. 
Quiet periods can be identified as periods between times when 
$|dS(t)/dt|$ exceeds a given threshold. However, the
durations of the quiet periods so defined
clearly depend on the magnitude of the
threshold (discussed further below) 
and the functional form of the wings. Histograms
representing the probability densities for
the durations of quiet periods for both models are shown
together in Fig.~\ref{fig:dur}(a). Both histograms display approximate
power-law behavior over at least five decades, but, somewhat
surprisingly, the powers are significantly different: 
approximately 1/time$^2$ for Model A
and 1/time for Model B. In contrast, the probability densities of the
lifetimes of individual species (the time from the population of a
species becomes nonzero until it again vanishes) are very close for the
two models, both closely following a 1/time$^2$ power law over more than
six decades (see Fig.~\ref{fig:dur}(b)). I believe this discrepancy
between the distributions of the durations of quiet periods and the
lifetimes of individual species is related to the fact that extinction
events are much more highly synchronized in Model A than in Model B. 
This can be clearly seen from the time evolution of the community
structure of the two models, shown in Fig.~\ref{fig:spec}. 
In Model A all the major species in a community usually go extinct
within a relatively short period of time, linking the duration of a quiet
period closely to the lifetimes of its major constituent species. As a
result, both the lifetime and quiet-period distributions for Model A
show approximate 1/time$^2$ behavior. Conversely, extinctions in Model B
tend to trigger further extinctions of only a part of the total food web
that describes the community, producing a much weaker synchronization 
between individual extinctions and the duration of relatively quiet
periods. This ability of the food web to avoid global
collapse triggered by the extinction of a single species 
may be the cause of the wide 1/time distribution of quiet
periods in this model. However, the 1/time distribution does not appear
to be a direct consequence of the low connectance of the 
interactions for this model: tests of Model B 
with a connectance of 100\% and 10\% potential producers also resulted
in a duration distribution with approximate 1/time decay. 

The time series for $|dS(t)/dt|$ is highly intermittent, and while the
quiet periods are power-law distributed in both models, individual
active periods, during which the activity remains continuously above the
cutoff threshold, are generally exponentially distributed with a mean
of just a few generations, except for very small thresholds. 
This was discussed for Model A in Ref.~\onlinecite{RIKV03}, and I
find the situation to be the same for Model B. 

\begin{figure}[t]
\begin{center}
\begin{tabular}{c}
\includegraphics[angle=0,width=.45\textwidth]{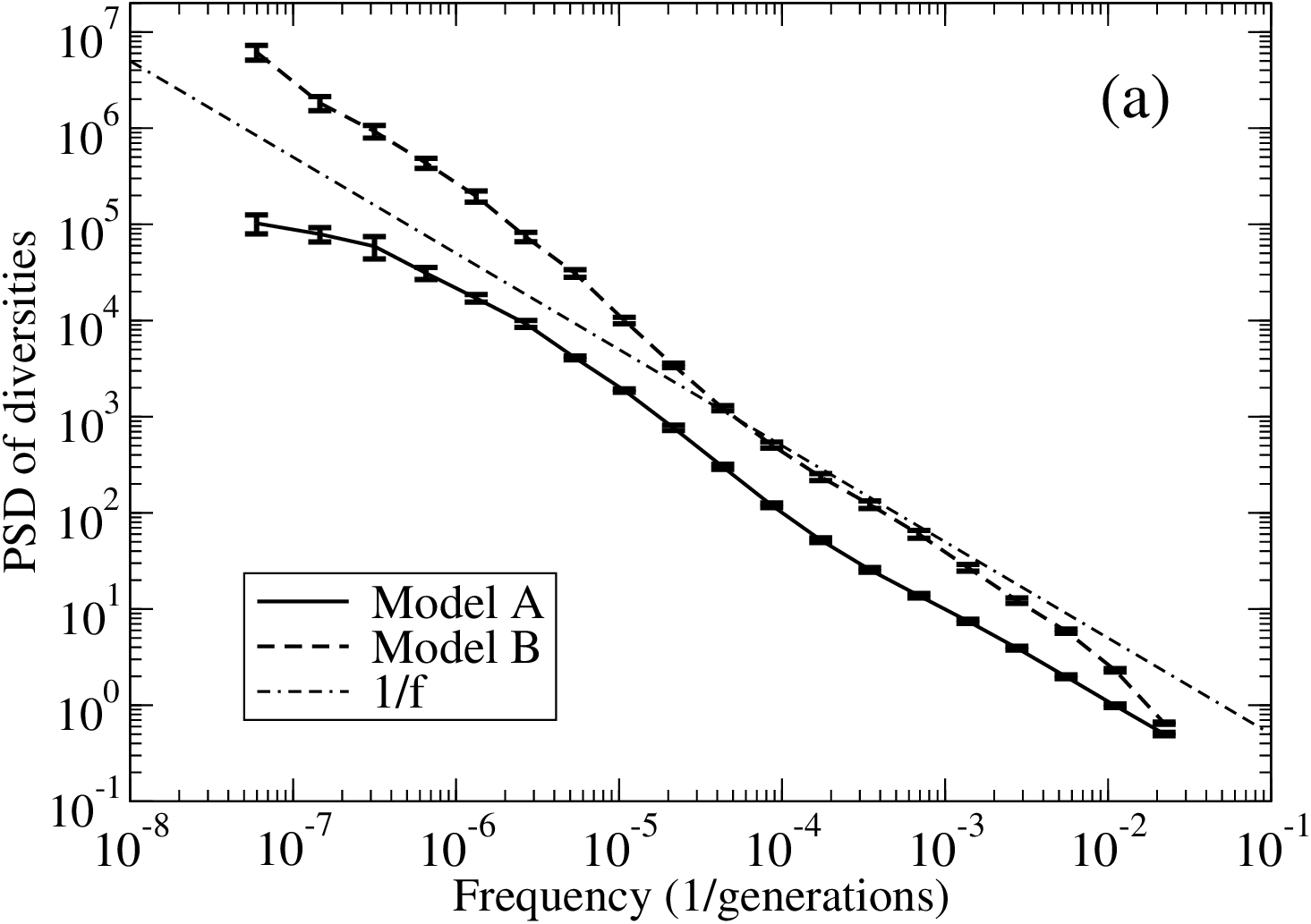}
\hspace{0.5truecm}
\includegraphics[angle=0,width=.45\textwidth]{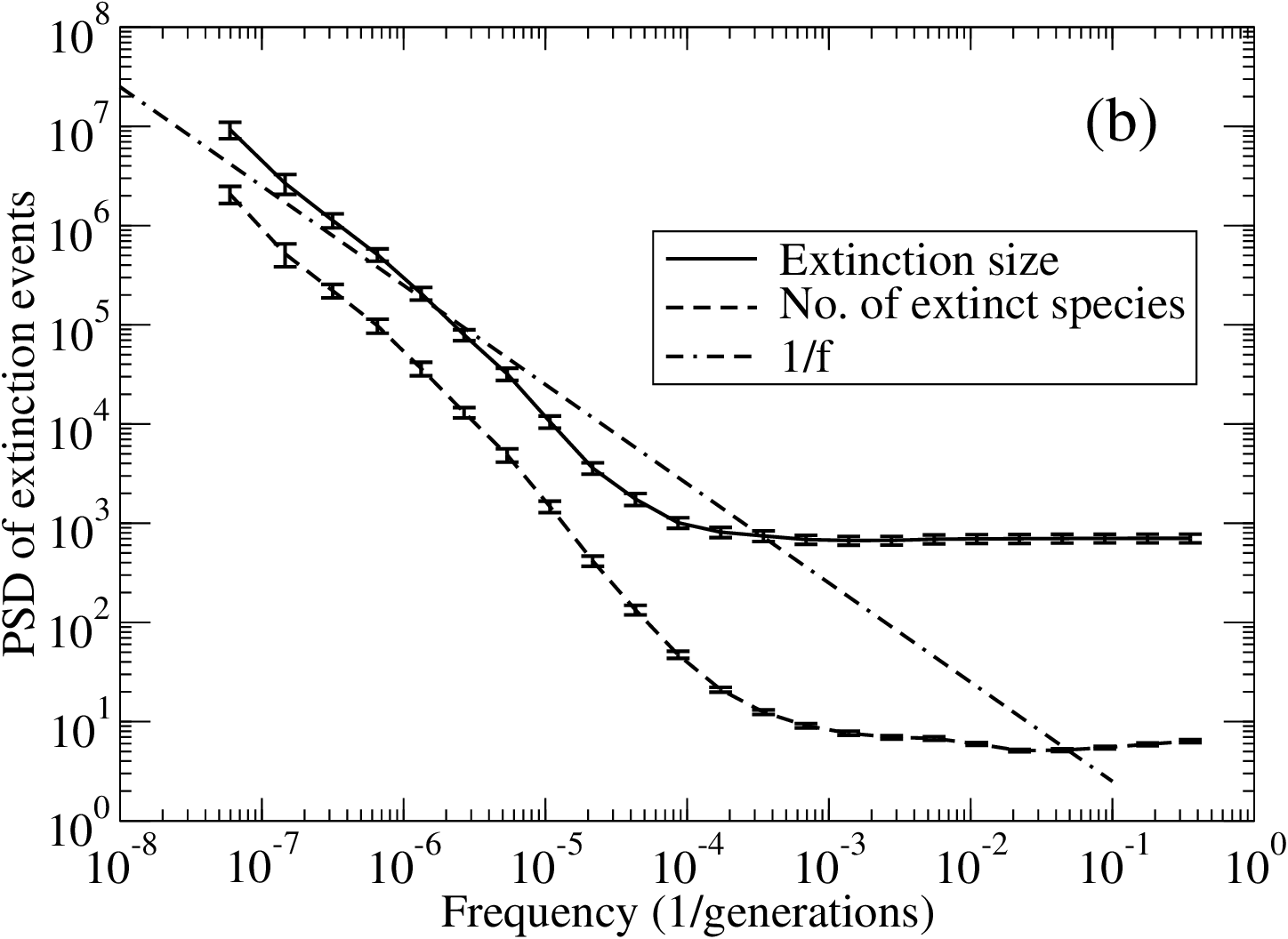}
\end{tabular}
\end{center}
\caption[example]
{\label{fig:PSD}
{\bf (a):}
Log-log plot of 
octave-averaged power spectral densities (PSD) for the diversity 
$D(t)$ for Model A
(solid) and Model B (dashed). The straight, dot-dashed line is a
guide to the eye, corresponding to $1/f$ behavior. The data were
averaged over 16 independent runs for Model A and 12 runs for Model B,
and the error bars are standard errors based on the differences between
independent runs. 
{\bf (b):}
Log-log plot of octave-averaged PSDs for the number of species going
extinct in one generation (dashed) and for the number of
species going extinct, weighted by the maximum population attained by each 
species, or {\it extinction size\/} (solid). 
Both PSDs refer to Model B and are
averages over 12 independent runs. The straight, 
dot-dashed line corresponds
to $1/f$ behavior, and error bars were calculated as in part (a). 
}
\end{figure}
An analysis method that is less
dependent on the details of the distribution of the large fluctuations
and the choice of cutoff,
is the power spectral density (PSD) of a time series of a suitable
global activity measure.
It may be reasonable to consider both models as examples of {\it extremal
dynamics\/}, in which the least ``stable" parts of the system
``collapse," leading to avalanches of readjustments throughout the
system. It is known for such systems\cite{PACZ96}
that if the probability density of the times between these local events 
has the long-time power-law dependence $\sim 1/t^{\tau_1}$ with 
$\tau_1 \le 2$, then the PSD of the global activity will depend
on frequency as $1/f^\alpha$, where $\alpha = \tau_1 -1$ (with a hard-to
detect logarithmic correction if $\tau_1=2$).\cite{PROC83}
In the systems considered here, it seems reasonable to 
identify these local waiting times
with the lifetimes of individual species, such that 
$\tau_1 \approx 2$ for both models. This reasoning then predicts that
the PSD of the diversity in both cases should exhibit practically 
pure $1/f$ noise (i.e., $\alpha \approx 1$). 
Indeed, the PSDs for both models, which are shown in Fig.~\ref{fig:PSD}(a),
exhibit $1/f$ like behavior over many decades. Except for a difference
in their overall magnitudes, the two PSDs behave essentially
alike for frequencies between $10^{-6}$ and 10$^{-2}$ generations$^{-1}$. 
For the very
lowest frequencies, the PSD for Model A appears to level off, but the
amount of data in this regime is small, and much longer simulations
would be needed to accurately determine the PSDs below 
$10^{-6}$ generations$^{-1}$. Power-law distributions and 
PSDs that fall off as $1/f^\alpha$ with $\alpha \approx 1$ have
been obtained from the fossil record.\cite{SOLE97} 
However, such power-law behaviors have only been observed over one or
two decades in time or frequency, and there are arguments to the
effect that $1/f$ spectra extracted in Ref.~\onlinecite{SOLE97} may be
severely influenced by the analysis method.\cite{NEWM99} 

Another quantity that can be used to determine the severity of an
extinction event is the total number of species that go extinct in a
single generation. However, this quantity is heavily influenced by the
repeated extinctions of small populations of unsuccessful mutants of
species with large populations. A better measure might therefore be
the number of species that go extinct, weighted by the maximum population
attained by each species during its lifetime. For convenience I shall
call this latter measure the {\it extinction size\/}. It can be seen as
a rough measure of the maximum fitness attained by the species that go
extinct in a particular generation. PSDs of both 
quantities for Model B are shown in Fig.~\ref{fig:PSD}(b). Like the
diversity PSDs, both these PSDs show $1/f$ like behavior for low
frequencies. However, they also have 
a wide region of near-white noise for higher frequencies. 

In models governed by extremal dynamics, the exponents $\alpha$ for
the frequency dependence of the PSD, $\tau_1$ for the duration of
local events (here interpreted as the lifetimes of individual
species), and $\tau$ for the duration of quiet periods where the
activity measure (here $|dS(t)/dt|$) falls below some cutoff
$y_c$, are connected by the following scaling 
relations:\cite{PACZ96}
\begin{eqnarray}
\alpha &=& 1 - \mu \nonumber\\
\tau_1 &=& 2 - \mu \nonumber\\
\tau   &=& 1 + \mu - \sigma
\;,
\label{fig:expo}
\end{eqnarray}
where $\mu$ and $\sigma$ are two independent exponents that depend on
the spatial dimensionality $d$ of the underlying model and both vanish 
in the limit $d \rightarrow 0$.\cite{DORO00} One thus notes that
(assuming that my identification of the quantities measured in the
models studied here with quantities in the extremal-dynamics
systems is valid) 
the exponents $\alpha=1$, $\tau_1=2$, and $\tau=1$, observed
for Model B, coincide with those of a zero-dimensional
extremal-dynamics system. On the other hand, Model A does not seem
to fit into this pattern.
It remains a question for further study to decide how appropriate
this analysis is. 

\begin{figure}[t]
\begin{center}
\begin{tabular}{c}
\includegraphics[angle=0,width=.45\textwidth]{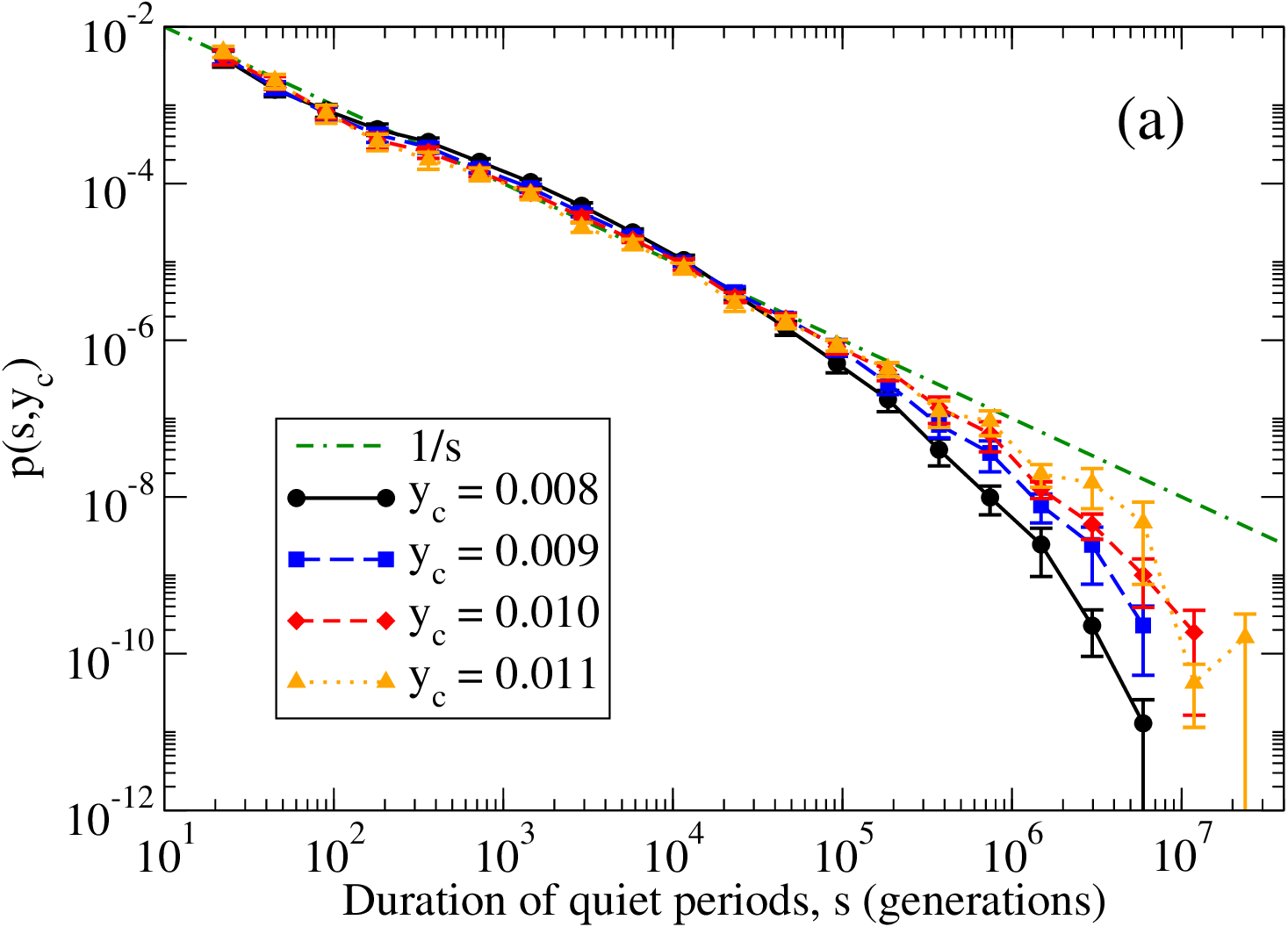}
\hspace{0.5truecm}
\includegraphics[angle=0,width=.45\textwidth]{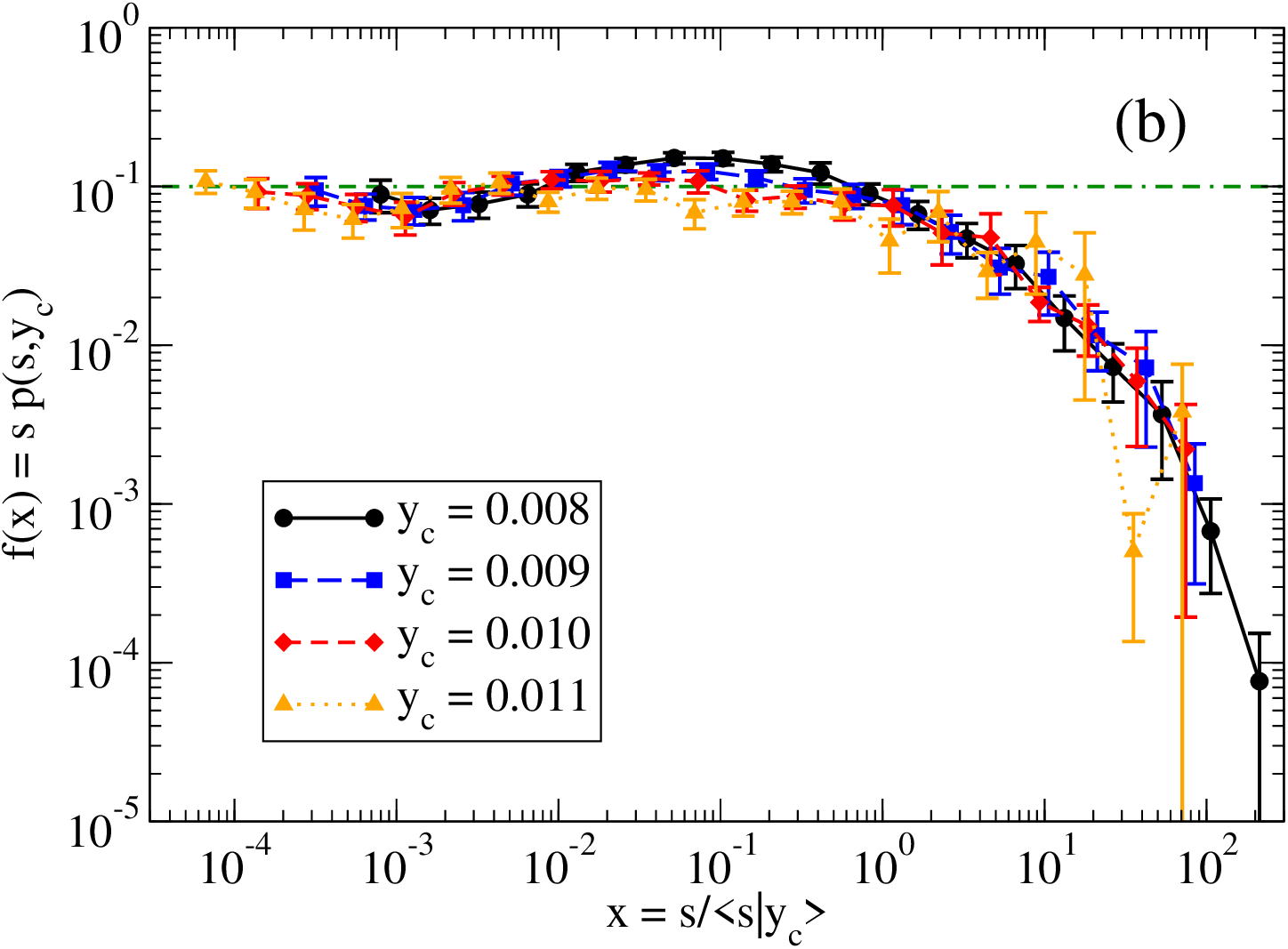}
\end{tabular}
\end{center}
\caption[example]
{\label{fig:durscale}
{\bf (a):}
Log-log plots of histograms representing the probability density 
$p(s,y_c)$
of the duration $s$ of quiet periods for Model B, as obtained with
different values of the cutoff $y_c$. 
The dot-dashed straight line corresponds to $1/s$ power-law behavior. 
{\bf (b):}
Log-log plot of the scaling function $f(x) = s p(s,y_c)$ vs 
the scaling variable $x = s/\langle s | y_c \rangle$. 
}
\end{figure}

As seen in Fig.~\ref{fig:dur}(a), the probability density of the
duration $s$ of quiet periods in Model B is well described as 
$p(s) \sim s^{- \tau}$ with $\tau \approx 1$, multiplied by a
function that approaches zero for large $s$ and a constant for
small $s$. 
Figure~\ref{fig:durscale}(a) shows that this function depends on
the cutoff $y_c$, and it is naturally expressed as a
scaling function, 
\begin{equation}
p(s,y_c) = s^{-\tau} f(s/g(y_c))
\;.
\label{eq:scale}
\end{equation}
The mean value of $s$ as calculated from the numerical $p(s,y_c)$,
$\langle s | y_c \rangle$, would seem a natural choice for $g(y_c)$, as 
long as it is much less than the total time $T$ of the simulation. 
Thus, the argument of the scaling function would be $x = s/\langle
s | y_c \rangle$.  In the relatively narrow range of
the four values of $y_c$ used in Fig.~\ref{fig:durscale}(a), 
$\langle s | y_c \rangle$ is well approximated by the power law 
$\langle s | y_c \rangle = 27948 \, (y_c/0.008)^{\gamma}$ with $\gamma 
\approx 7.84$, and the scaling function, 
\begin{equation}
f(x) = s^\tau p(s,y_c) = x^\tau \langle s | y_c \rangle^{\tau-1} 
       \langle s | y_c \rangle p(x \langle s | y_c \rangle, y_c)
     \equiv 
       x^\tau \langle s | y_c \rangle^{\tau-1} \tilde{p}(x) 
\;.
\label{eq:scale2}
\end{equation}
The resulting $f(x)$ is shown in Fig.~\ref{fig:durscale}(b) for
$\tau=1$. The data collapse is reasonable, but it is not very
sensitive to the value of $\gamma$, which can be reduced to near 5
without giving a significantly inferior collapse. 

\section{CONCLUSION}
\label{sec:conc}  

In this paper I have discussed the fluctuations in two different,
simplified models of biological macroevolution that both are based
on the birth/death behavior of single individuals on the ecological
timescale. On evolutionary timescales both models give rise to power-law
distributions of characteristic waiting times and power spectra
that exhibit $1/f$ like flicker noise, in agreement with some
interpretations of the fossil record.\cite{SOLE97,NEWM99}
The main difference in the
construction of the models is that in Model A the population size
is controlled by a Verhulst factor, while in Model B the population
is maintained by a small percentage of the species that have the
ability to directly utilize an external resource (producers or
autotrophs). All other species must maintain themselves as
predators (consumers or heterotrophs). In both models the
probability density of the lifetimes of individual species follows a 
1/time$^2$ power law and, consistently, the power spectra of the
time series of diversity and extinction sizes show $1/f$ noise.
However, the probability density of quiet periods, defined as the
times between events when the magnitude of the logarithmic derivative 
of the diversity exceeds a given cutoff, behaves differently in the
two models. In Model B, the quiet-period distribution goes as
1/time, consistent with the universality class of the
zero-dimensional Bak-Sneppen model. In contrast, in Model A the behavior
is proportional to 1/time$^2$, like the lifetimes of individual species. 
The difference between the behaviors can be linked to the lower
degree of synchronization between extinction events in Model B. 
It remains a topic for further research to see whether this
difference extends to more realistic modifications of Model B. 

\acknowledgments     

I thank R.~K.~P.\ Zia for a pleasant and fruitful 
collaboration in formulating and studying Model A, 
V.~Sevim for the data on Model A that appear in
Fig.~\ref{fig:dur}(b), and J.~W.\ Lee for useful discussions. 

This work was supported in part by the U.S.\ National Science
Foundation through Grants No.\ DMR-0240078 and DMR-0444051, and by 
Florida State University through the School of Computational
Science, the Center for Materials Research and Technology, and the
National High Magnetic Field Laboratory. 



\end{document}